\documentclass{PoS}

\usepackage{amsmath,amssymb,mathtools}
\usepackage{enumerate}
\usepackage{booktabs}
\usepackage{multirow}
\usepackage{graphicx}
\usepackage{xspace}
\usepackage{subcaption}

\title{Scale setting in inclusive jet production at hadron colliders}

\ShortTitle{Scale setting in inclusive jet production at hadron colliders}

\author{\speaker{Jo\~{a}o Pires}\\
        Centro de F\'{\i}sica Te\'{o}rica de Part\'{\i}culas - CFTP, 
Instituto Superior T\'ecnico IST,
Universidade de Lisboa, Av.\ Rovisco Pais,
P-1049-001 Lisboa, Portugal\\
        E-mail: \email{joao.ramalho.pires@tecnico.ulisboa.pt}}

\abstract{In this contribution we review the status of inclusive jet cross sections at hadron colliders in perturbative QCD, studying in particular the
renormalisation and factorisation scale dependence of the one-jet inclusive cross section at NNLO.}

\FullConference{Sixth Annual Conference on Large Hadron Collider Physics (LHCP2018)\\
                4-9 June 2018\\
                Bologna, Italy}

\begin{document}

\section{Introduction}
The theoretical framework for the study of inclusive jet production at hadron colliders is the improved parton model formula where the inclusive cross section for the hard scattering
process initiated by two hadrons with momenta $P_{1}$ and $P_{2}$ can be written as, 
\begin{equation}
  \label{eq:hadroncross}
  {\rm d}\sigma(P_1,P_2) = 
  \sum_{i,j} \int_0^1{\rm d} x_1 \int_0^1{\rm d} x_2 \; f_i(x_1,\mu_F) \; f_j(x_2,\mu_F)
  \;{\rm d}\hat{\sigma}_{ij}(\alpha_s(\mu_R),\mu_R,\mu_F,x_1P_{1},x_{2}P_{2})\;.
\end{equation} 
For sufficiently high energies $Q^2 \gg\Lambda_{QCD}$, the initial state partons behave as free particles, such that the approximate factorized
form of the cross section is justified. In~\eqref{eq:hadroncross}, the parton distribution functions $f_{i}(x,\mu_F)$ are the usual quark and gluon distribution functions defined at a factorization 
scale $\mu_{F}$, while the short-distance cross section for the scattering of partons $i$ and $j$ is denoted by $\hat{\sigma}_{ij}$. Furthemore, for sufficiently high $p_{T}$ jet production, the scale 
of the reaction is such that the strong coupling constant decreases and the partonic cross section can be computed in perturbation theory from first principles from the Lagrangian of QCD,
\begin{equation}
\hat{\sigma}_{ij}=\sum_{m=2}^{n}c_{m}\alpha_{s}^{m}\;.
\end{equation}  

Much progress has been achieved on the theory side towards obtaining fixed-order and resummed calculations for inclusive jet production at hadron colliders and in matching the accuracy of the perturbative calculation 
with parton showers.
First order (NLO) corrections in the strong coupling constant were obtained for first time in~\cite{Ellis:1992en,Giele:1994gf,Nagy:2001fj} while NLO corrections in the electroweak theory were
obtained in~\cite{Dittmaier:2012kx,Campbell:2016dks,Frederix:2016ost}. Subsequently, the matching of the NLO QCD prediction to the parton shower was obtained in the frameworks of POWHEG~\cite{Alioli:2010xa} and MC@NLO~\cite{Hoche:2012wh}. 

More recently, there has been progress in developing factorization theorems for single jet inclusive production to allow the joint of resummation of threshold logarithms~\ $\alpha_s^n\left(\ln^k(z)/z\right)_+$, where $z$ is the invariant mass of the system recoiling against the jet~\cite{Liu:2017pbb}, in combination with the resummation of small-$R$ size jets $\alpha_{s}^n\log^n(R)$~\cite{Dasgupta:2016bnd,Liu:2017pbb}. The hard perturbative functions in the resummed cross section are know to NLO and allow the joint
resummation of the two logarithmic contributions at NLL+NLO accuracy via an additive matching to the fixed-order NLO QCD result~\cite{Liu:2017pbb}. In this case, the resummation scales can be evolved through RGE equations to a hard scale $\mu=p_{T}$ which is the $p_{T}$ of the jet in the $p_{T}$ and rapidity slice $y$ where it is observed. 

The latest theoretical development in the description of jet production at a hadron collider is the calculation of the
second order (NNLO) QCD corrections to the single jet inclusive cross section~\cite{Currie:2016bfm}. In this approach, the exact matrix elements of QCD that contribute at this order are included in a parton-level generator framework NNLOJET~\cite{Gehrmann:2018szu}, keeping the leading colour $N_c$ correction of the NNLO contribution. Remaining colour-suppressed contributions at this order are not yet available for all partonic channels,
however, it was observed that in the NLO full-colour cross section expansion, the subleading colour channels are already below the two percent level. In this framework, the analytic cancellation of infrared singularities at NNLO between the real-emission contribution and the virtual corrections is achieved using the antenna subtraction method~\cite{GehrmannDeRidder:2005cm,Currie:2013vh,Glover:2010im,GehrmannDeRidder:2011aa,Ridder:2012dg}. 

\begin{figure}[t]
  \centering
   \includegraphics[width=12cm]{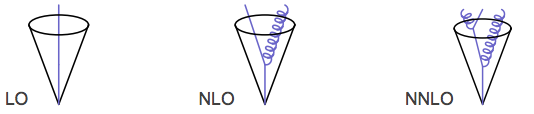}
  \caption{Illustrations of the description of the internal jet shape at LO, NLO and NNLO with cones representing jets with quark and gluon radiation inside the jet depicted in blue lines.  }
  \label{fig:jetshape}
\end{figure}

At this level of accuracy, we can observe that the NNLO prediction allows for a systematic reduction of the perturbative uncertainty of the result, from the inclusion of higher order terms in the perturbative expansion, producing a better physical description of the the hard scattering process. As an example, the renormalization scale  $\mu_{R}$-dependence of the inclusive jet prediction up to NNLO is given by,
\begin{eqnarray}
\sigma(\mu_R,\alpha_s(\mu_R),L_{R})&=&\left(\frac{\alpha_s(\mu_R)}{2\pi}\right)^2\sigma^{0}_{ij}+\left(\frac{\alpha_s(\mu_R)}{2\pi}\right)^3\left( \sigma_{ij}^{1} +2\beta_{0}L_R\sigma_{ij}^{0}\right)\nonumber\\
&+&\left(\frac{\alpha_s(\mu_R)}{2\pi}\right)^4\left( \sigma_{ij}^{2} +L_{R}(3\beta_0\sigma_{ij}^{1}+2\beta_1\sigma_{ij}^{0})+L_R^2 3\beta_{0}^2\sigma_{ij}^{0}\right)+{\cal O}(\alpha_s^5)\;,
\label{eq:muR}
\end{eqnarray}
where the NNLO correction $\sigma_{ij}^{2}$ is computed exactly from first principles and includes the scale compensation terms $L_R=\log(\mu_{R}/\mu_o)$ that reduce the scale dependence of the prediction at NLO. Moreover, at this order the final
state jets are modelled by extra partons as shown in Fig.~\ref{fig:jetshape} using the exact matrix elements such that perturbation theory starts to reconstruct the shower and its effects without approximations. We can observe at NNLO a better matching of the jet algorithm between the theory calculation and experimental setup enabling an improved understanding of the jet shape. 

\begin{figure}[t]
   \centering
   \begin{subfigure}{.5\textwidth}
    \centering
  \includegraphics[width=6cm]{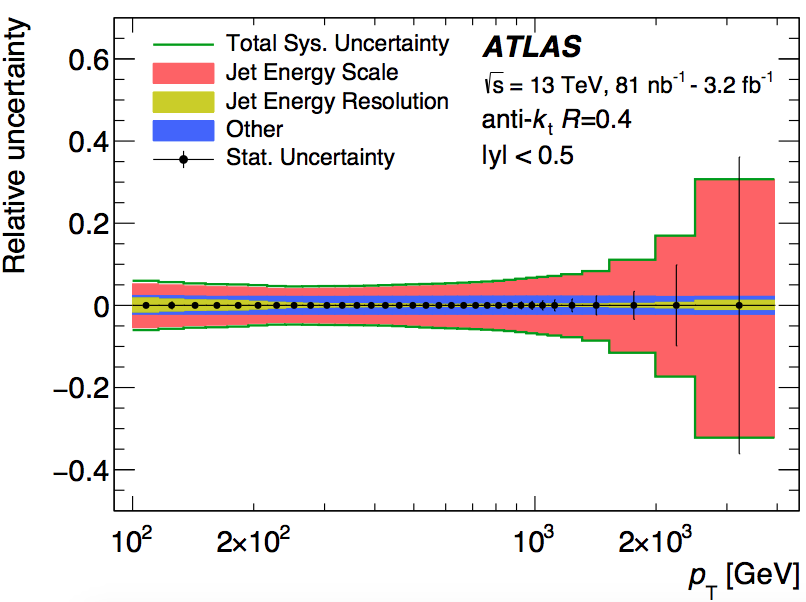}
   \caption{}
   \end{subfigure}%
   \begin{subfigure}{.5\textwidth}
     \centering
    \includegraphics[width=6cm]{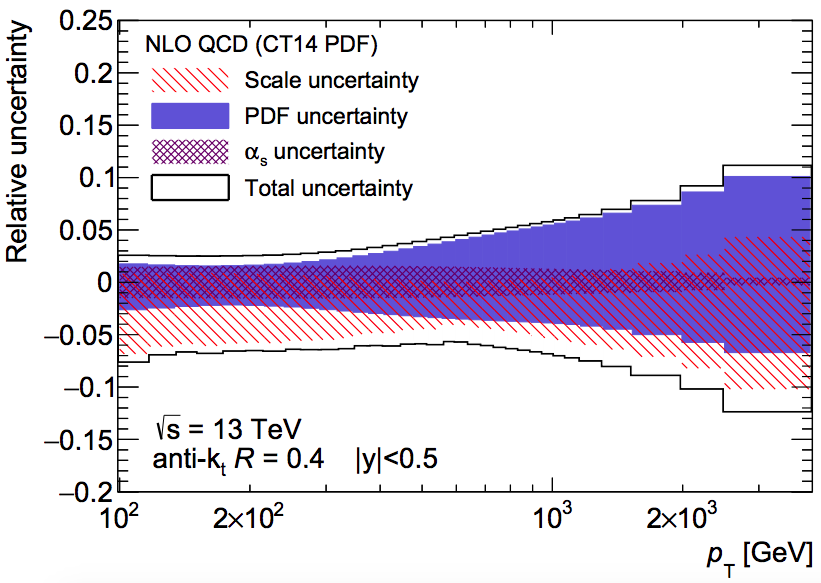}
   \caption{}
   \end{subfigure}%
   \caption{(a) Total and break-down of experimental systematic uncertainties for inclusive jet production at ATLAS at $\sqrt{s}$=13 TeV and (b), theoretical uncertainties at NLO coming from the scale uncertainty in the prediction, and the parametric uncertainty dependence on the parton distribution functions and $\alpha_s$~\cite{Aaboud:2017wsi}. }
  \label{fig:jetuncertainties}
\end{figure}
For these reasons the NNLO result provides the first serious estimate of the theory error allowing a comparison with the wealth of experimental data which have similar precision. As shown in Fig.~\ref{fig:jetuncertainties}, typical uncertainties on the experimental side are at the 5-10\% level dominated by the jet energy scale uncertainty. On the theory side, the scale uncertainty of the NLO result is at 10\% level, indicating that the NNLO prediction is needed to improve our understanding of the inclusive jet data produced at the LHC.

In this talk I will therefore concentrate on reviewing the latest work on the understanding of the perturbative uncertainty of the NNLO result beginning by identifying the most appropriate scale choices to study this process
and analysing the behaviour of their respective perturbative expansions.

\section{Inclusive jets scale choices}
Despite being unphysical parameters, a dependence on the renormalization and factorization scales remains in the cross section truncated at a given fixed order in the perturbative expansion. The dependence of the cross section on their values is absent for an all order calculation, and for this reason, their choice is completely and \emph{a priori} arbitrary. In the present section we will consider the following scale choices for the
process of inclusive jet production,
\begin{enumerate}[(a)]
\item $\mu=p_{T}$, when we use the $p_{T}$ of the jet in the $p_{T}$ and rapidity slice $|y|$ where the jet is observed as the scale to compute its contribution to the cross section.
\item $\mu=p_{T1}$ when we use  the $p_{T}$ of the hardest jet in the event as the scale of the event and for all jets in the event.
\item $\mu=H_{T}=\sum_{i,\in jets}p_{T,i} $ when we use the scalar sum of the $p_{T}$ of all jets in the event as the scale for the event.~\footnote{Despite being a natural scale choice, it is clear from its definition that this scale
suffers from a discontinuous behaviour when the number of reconstructed jets changes, leading to large displacements in cross section near the phase space boundaries where $n_{jets}\to n_{jets}+1$. As a consequence, higher order corrections are unstable with this scale choice and we will no longer consider this scale choice.}
\item $\mu=\hat{H}_{T}=\sum_{i,\in partons}p_{T,i} $ when we use the scalar sum of the $p_{T}$ of all partons in the event as the scale for the event.
\end{enumerate}
In order to assess the scale uncertainty of our results we will produce independent variations of each scale by factors of 2 
with the constraint $1/2\leq\mu_{R}/\mu_{F}\leq2$, to produce a scale uncertainty band for each perturbative QCD prediction.

\section{Results}
\begin{figure}[t!]
\centering
\begin{subfigure}{.5\textwidth}
  \centering
  \includegraphics[width=\linewidth]{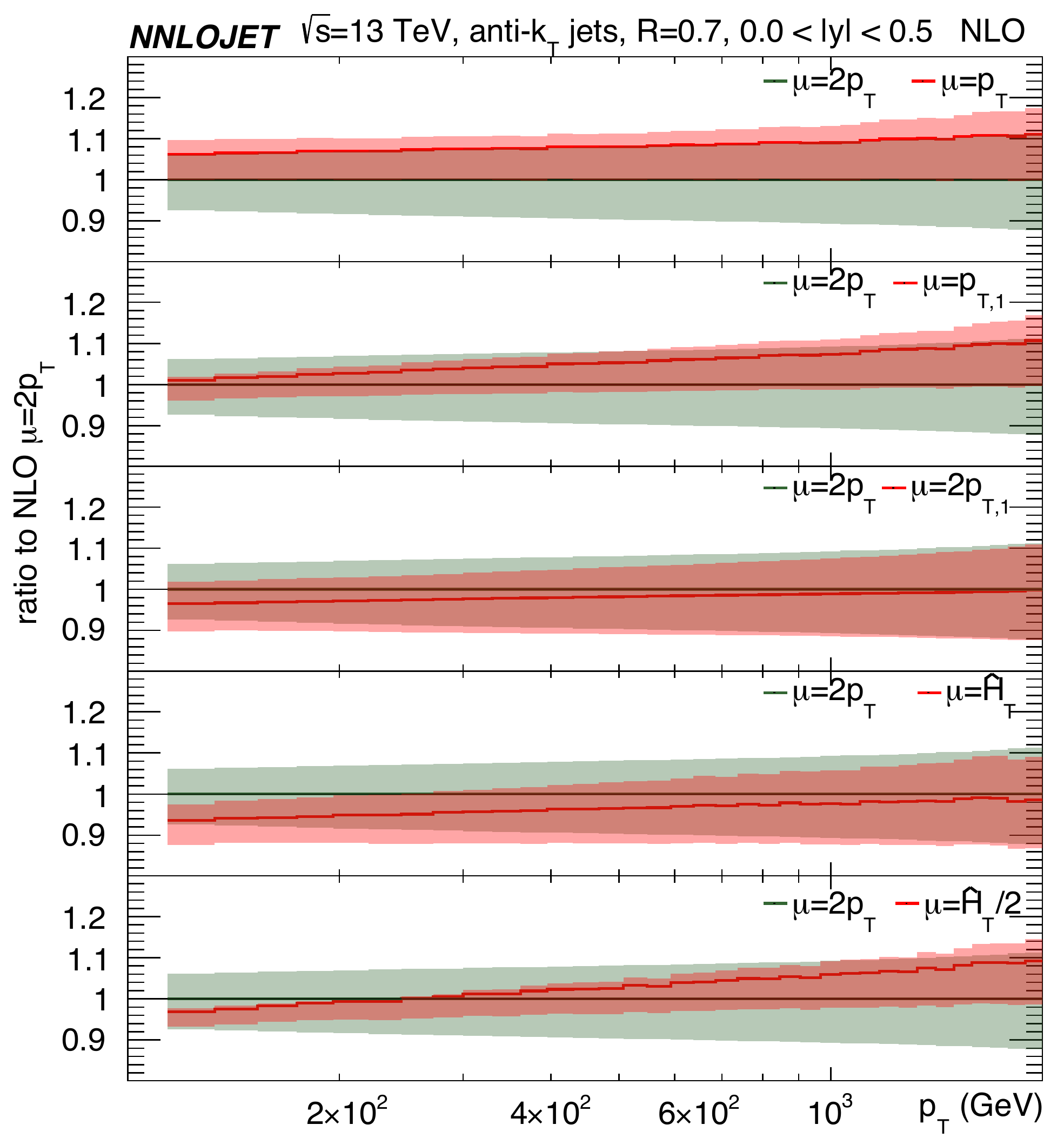}
  \caption{}
\end{subfigure}%
\begin{subfigure}{.5\textwidth}
  \centering
  \includegraphics[width=\linewidth]{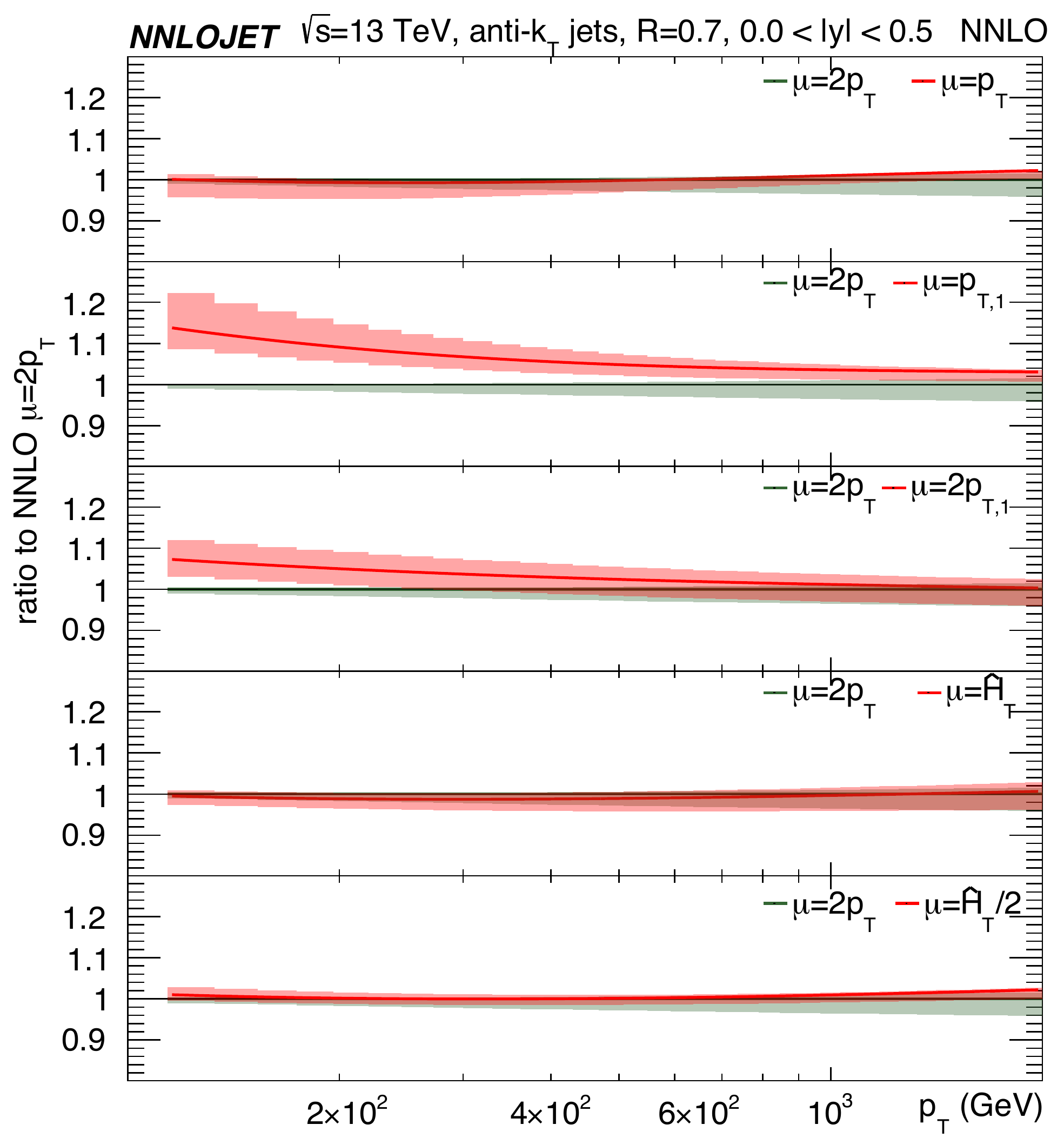}
  \caption{}
\end{subfigure}
\caption{Ratio of 13 TeV single jet inclusive cross sections to the $\mu=2\,p_{T}$ scale choice at (a) NLO and (b) NNLO with $R=0.7$ and CMS cuts~\cite{Currie:2018xkj}.}
\label{fig:ratiotoPT107}
\end{figure}
In Fig.~\ref{fig:ratiotoPT107} we present perturbative QCD predictions at NLO (on the left) and NNLO (on the right) for 6 different central scale choices for inclusive jet production at the LHC at a center of 
mass energy $\sqrt{s}=13$~TeV. Five scale choices identified in the previous section are labelled in the legend of both plots and plotted in red, normalised to the $\mu=2\,p_{T}$ scale choice in green. 
The shaded bands assess the scale uncertainty of each central scale choice and were obtained as was described in the previous section.

We can 
observe a dramatic reduction in the scale uncertainty when going from NLO to NNLO as anticipated by Eq.~\eqref{eq:muR}, as indicated by the reduction in the thickness of the red and green bands. Comparing now the 
behaviour of each individual central scale choice we observe a large spread in the predictions at NLO (on the left) which however is captured by the larger scale uncertainty of the NLO result, i.e., the red and green bands 
overlap. When comparing with the results at NNLO (on the right) we can observe at high $p_{T}$ an excellent agreement in the prediction of the cross section independently of the 
scale choice with the scale uncertainties at the few percent level. We observe however, larger 
differences at low $p_{T}$ where in the particular the scale choices $\mu=p_{T1}, \mu=2p_{T1}$ tend to look similar and predict a larger NNLO cross section of approximately 10\% with respect to the 
other scale choices. They also display a larger scale 
uncertainty band when compared with the other scale choices. For this reason, a spread in the NNLO predictions at the level of 10\% in the low $p_{T}$ region is still slightly larger than the target reach in precision of scale 
uncertainties at the percent
level for this observable. For this reason, it is the goal of the next section to compare the behaviour of the perturbative expansion of the NNLO predictions as a function of the central scale choice,
making use of the knowledge of three orders in the perturbative expansion of the observable, to understand the source of discrepancy between the $\mu=p_{T}$ and $\mu=p_{T1}$ type of scale choice.

\section{Comparison of different scale choices}
\begin{figure}[t]
\centering
\begin{subfigure}{.5\textwidth}
  \centering
  \includegraphics[width=\linewidth]{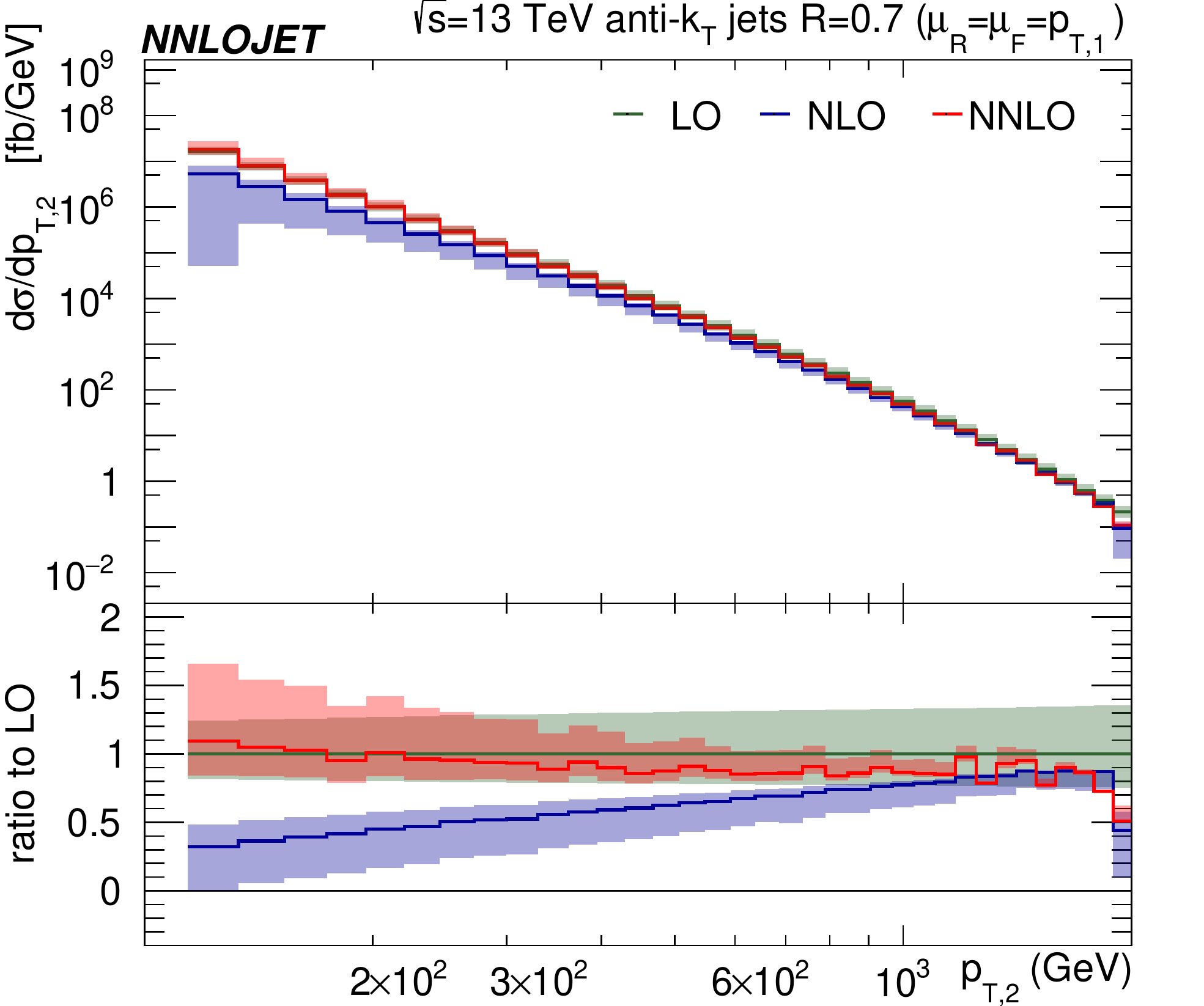}
  \caption{}
  \label{fig:PT2muPT1R07}
\end{subfigure}%
\begin{subfigure}{.5\textwidth}
  \centering
  \includegraphics[width=\linewidth]{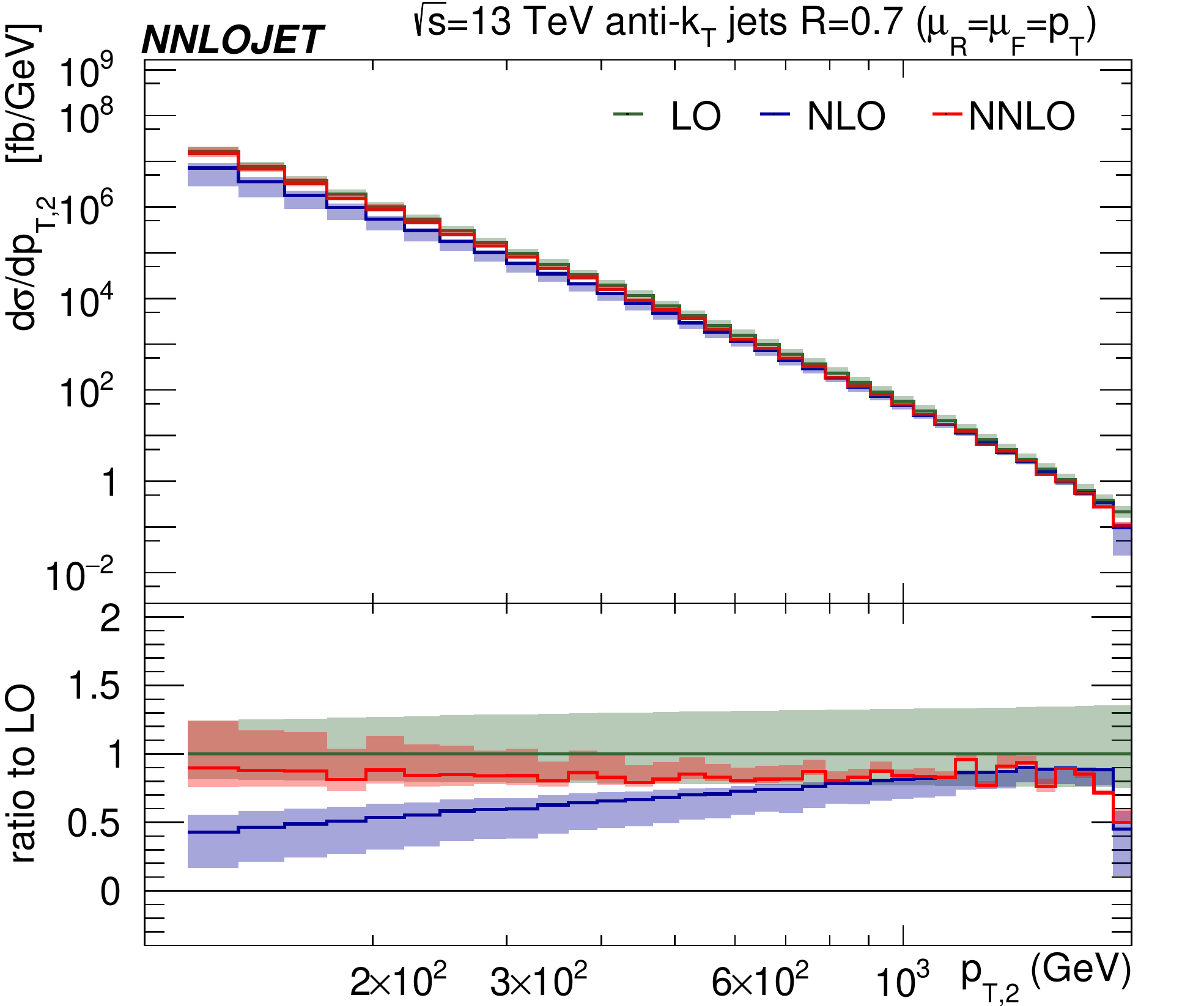}
  \caption{}
  \label{fig:PT2muPTR07}
\end{subfigure}
\caption{Perturbative corrections to the transverse momentum  distribution of the second jet 
at 13 TeV (CMS cuts, $|y| < 4.7$, $R=0.7$),  integrated over rapidity and normalised to 
the LO prediction. Central scale choice: (a) $\mu=p_{T,1}$ and (b) $\mu=p_{T}$. 
Shaded bands represent the theory uncertainty due to the variation of the factorization and renormalization scales~\cite{Currie:2018xkj}.}
\label{fig:PT2R07}
\end{figure}

As shown in~\cite{Currie:2018xkj}, the one-jet inclusive cross section can be decomposed into its individual subleading jet contributions through ${\cal O}(\alpha_s^4)$ as,
\begin{eqnarray}
\frac{{\rm d}\sigma}{{\rm d}p_{T}}(\mu=p_{T,1})&=&\frac{{\rm d}\sigma}{{\rm d}p_{T,1}}(\mu=p_{T,1})+\frac{{\rm d}\sigma}{{\rm d}p_{T,2}}(\mu=p_{T,1})+\frac{{\rm d}\sigma}{{\rm d}p_{T,3}}(\mu=p_{T,1})
+\frac{{\rm d}\sigma}{{\rm d}p_{T,4}}(\mu=p_{T,1})\;.\nonumber
\end{eqnarray}

However, when decomposing the inclusive jet cross section in terms of the contributions from leading and subleading jets, 
the individual jet distributions are well-defined and infrared-safe 
only if they are inclusive in the jet rapidity (with the same global rapidity 
cuts applied to all jets). 
Since the notion of leading and sub-leading jet is not well defined at leading order ($p_{T,1} = p_{T,2}$ at LO), the 
rapidity assignment to the leading and subleading jet is ambiguous for leading-order kinematics. When 
computing higher-order corrections, the rapidity of the leading and subleading jet may thus be interchanged between 
event and counter-event, 
causing them to end up in different rapidity bins, thereby obstructing their cancellation in infrared-divergent 
limits. On the other hand, in the inclusive jet transverse momentum distribution (which sums over all jets in the event)
IR-safety is restored in differential distributions in rapidity $y$, since leading and subleading jet contributions are 
treated equally. For this reason, this is the observable presented in the experimental measurement. 

For the observable at hand, we have observed in~\cite{Currie:2018xkj}, that the first and second jet constitute the dominant contributions to the inclusive jet sample, while the third and fourth jet rates 
are suppressed by
additional powers of $\alpha_s$ and are completely negligible. For this reason, we concentrate on the transverse momentum of the second jet contribution to the inclusive jet sample, and plot its transverse momentum
distribution as a function of the scale choice in Fig.~\ref{fig:PT2R07}.

\begin{figure}[t]
\centering
\begin{subfigure}{.5\textwidth}
  \centering
  \includegraphics[width=\linewidth]{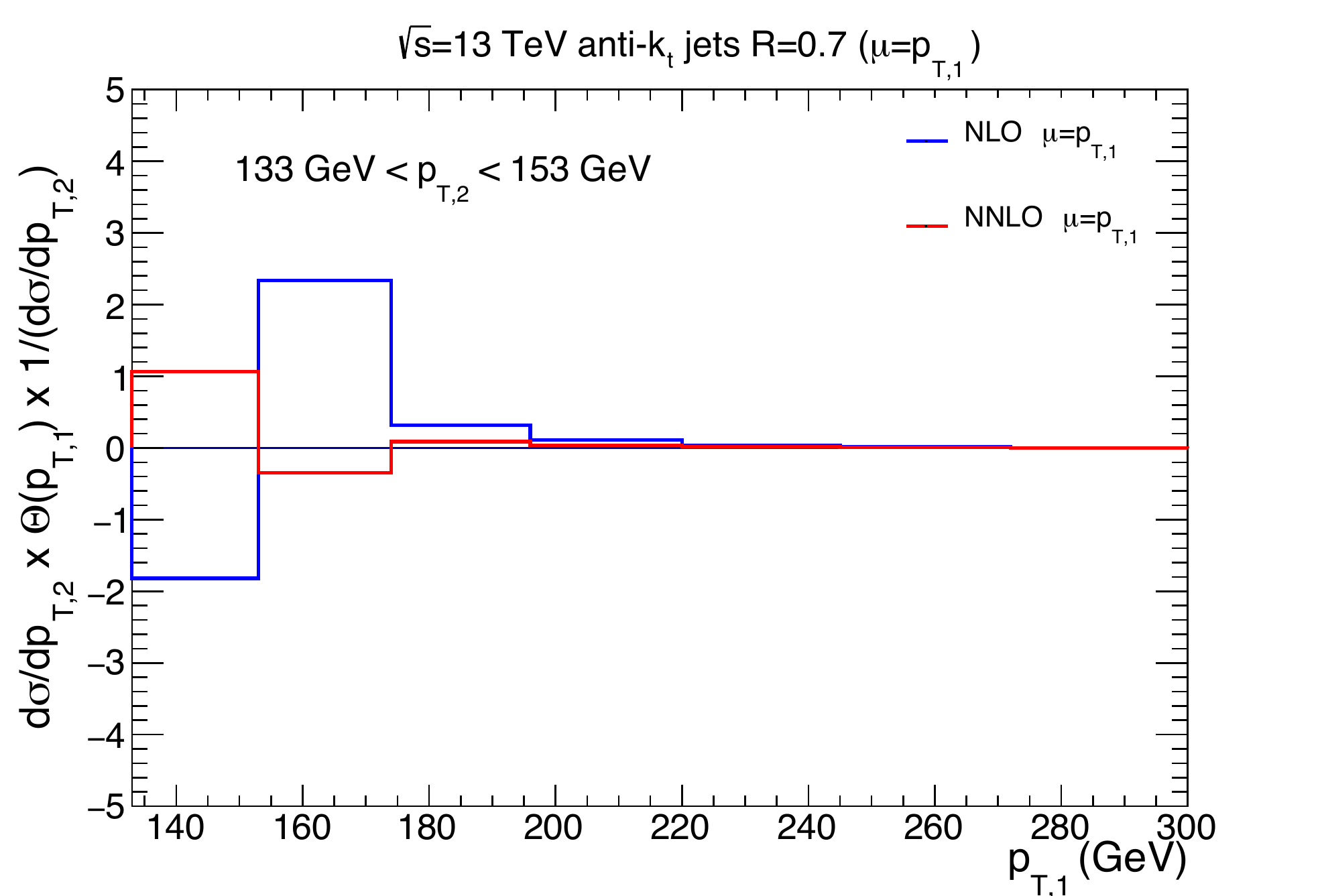}
  \caption{}
\end{subfigure}%
\begin{subfigure}{.5\textwidth}
  \centering
  \includegraphics[width=\linewidth]{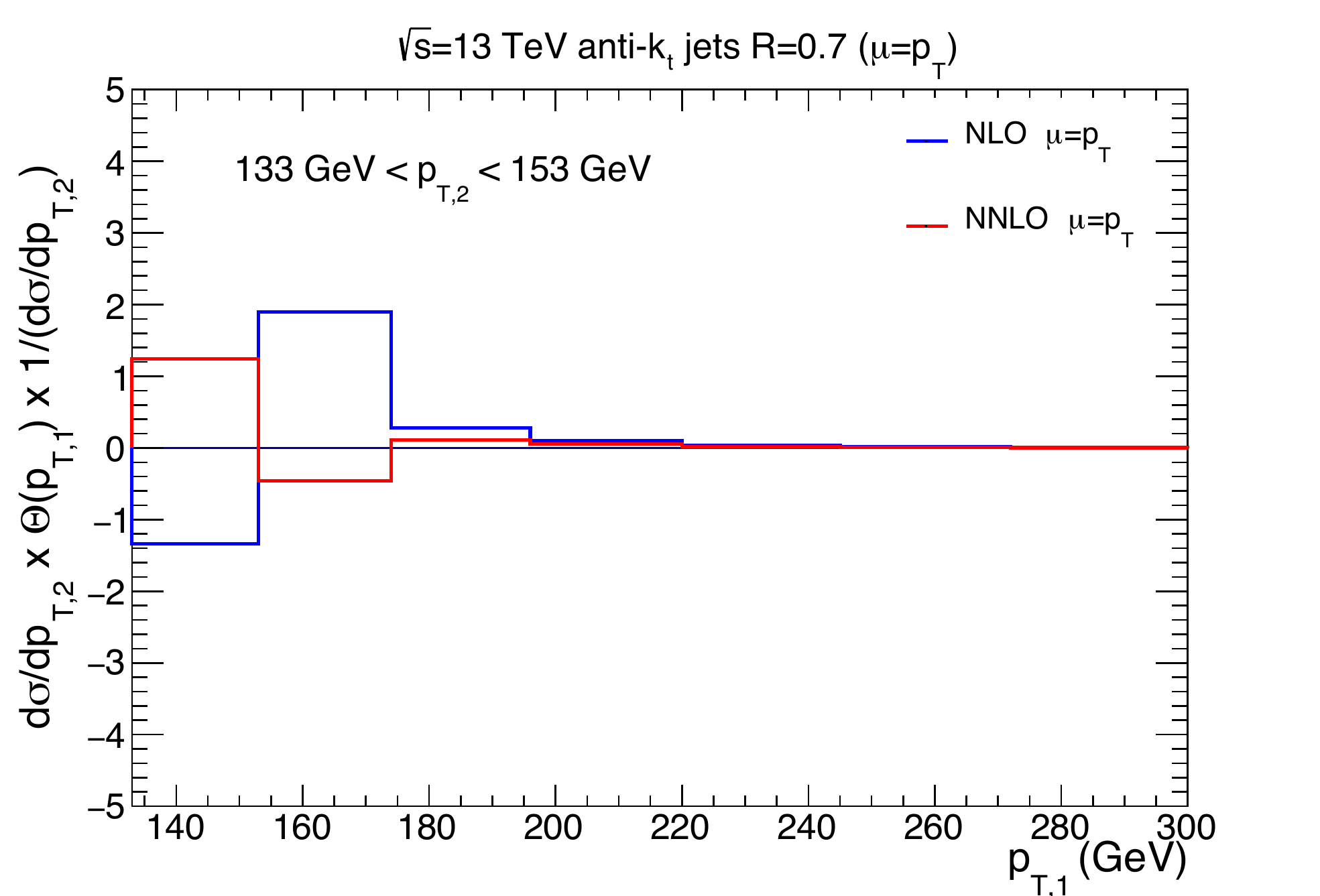}
  \caption{}
\end{subfigure}
\caption{Decomposition of events contributing to a single bin in $p_{T,2}$ according to the transverse momentum 
of the leading jet in the event $p_{T,1}$. CMS cuts at 13 TeV with jet resolution $R=0.7$ and scale choice (a) 
$\mu=p_{T,1}$; (b) $\mu =p_{T}$~\cite{Currie:2018xkj}.}
\label{fig:PTimb}
\end{figure}

By comparison of the two predictions (for $\mu=p_{T,1}$ on the left and $\mu=p_{T}$ on the right), we can conclude that this contribution is very sensitive to IR-effects and exhibits (for both scale choices), an
alternating series expansion with large coefficients. It is reassuring that both predictions are stabilised at NNLO, and in line with the LO result, however, the functional form of the scale choice has an impact. In particular,
we can observe smaller perturbative coefficients for the scale choice $\mu=p_{T}$ than for $\mu=p_{T,1}$, which points to a faster convergence of the perturbative expansion with the former scale choice. In order to observe
why this effect is improved for the $\mu=p_{T}$ scale choice we show in Fig.~\ref{fig:PTimb} the fractional contribution to the second jet $p_T$ distribution in a given $p_{T,2}$ 
interval (133 GeV < $p_{T,2}$ < 153 GeV) for particular $p_{T,1}$ slices plotted along the horizontal axis.

The bin content 
is constrained to sum to unity by construction. We observe 
that this is achieved from a large cancellation (for both scale choices) 
between the first bin of the distribution (where $p_{T,1}=p_{T,2}$) and the adjacent bin where ($p_{T,1}\gtrsim p_{T,2}$). In particular at NLO (in blue) the entire second bin
content is filled from the NLO real emission (where $p_{T,1}$ can be larger than $p_{T,2}$ for the first time) while the virtual correction contributes to the first bin only.
When comparing the behaviour of the two scale choices we note that for $\mu=p_{T,1}$ the scale is increasing along the $x$-axis.
On the other hand for $\mu=p_{T}$, the scale is fixed to be equal to $p_{T,2}$ for all contributions and the cancellation between the large positive real emission and large negative virtual
correction is improved (as shown by the height of the bins). This effect is even more pronounced for the $R=0.4$ jet size~\cite{Currie:2018xkj}. This is due to the fact that for the smaller jet cone size
we promote more events with relatively soft emissions, which are not recombined into outgoing jets and generate an imbalance between $p_{T,1}$ and $p_{T,2}$. In that case, the incomplete
cancellation at low $p_{T}$ for the large negative virtual correction and large positive real-emission contribution can be aggravated for certain types of scale choices as shown in~\cite{Currie:2018xkj}. 

This observation has motivated us to introduce an extended set of criteria to help identify the most appropriate scale choice for the perturbative description of single jet inclusive production at
hadron colliders~\cite{Currie:2018xkj}, which I review in the following section. 

\section{Scale choice criteria}
As was shown in the previous section, the inclusive $p_T$  distribution suffers from an infrared sensitivity that exhibits a strong dependence on the
scale that is used and a suboptimal choice can introduce pathological behaviours in the predictions.
In the following, I summarise an extended set of criteria introduced in~\cite{Currie:2018xkj} to help identify the most appropriate scale choice for the perturbative description of single jet inclusive production at
hadron colliders,

\begin{enumerate}[(a)]
\item perturbative convergence of the inclusive jet sample, i.e., the size of the perturbative corrections reduces in magnitude at each successive order in the perturbative expansion.
\item scale uncertainty as theory estimate, i.e, the scale uncertainty bands of the theory prediction should overlap between the last two orders, i.e., NLO and NNLO.
\item perturbative convergence of the individual jet spectra, i.e., the perturbative corrections to the leading and subleading jet decrease in magnitude at each successive order in the perturbative expansion.
\item stability of the second jet-$p_{T}$ distribution, i.e., we require the predictions and associated scale uncertainties to provide physical, positive cross sections.
\end{enumerate}

As documented in~\cite{Currie:2018xkj}, not all scale choices satisfy these criteria. In the Table~\ref{tab:scale-criteria} we summarise the behaviour of the various central scale choices for $R=0.4$ and $R=0.7$. 

We can conclude from Table~\ref{tab:scale-criteria}, that we can single out $\mu=2p_{T}$ and $\mu=\hat{H}_{T}$ as scale choices that satisfy all the criteria above for both cone sizes $R=0.4$ and $R=0.7$ while,
for example, the choice $\mu=p_{T,1}$ is
strongly disfavoured. In this way, we are able to eliminate certain scale choices that introduce pathological behaviours in the perturbative expansion of the observable and thereby remove certain ambiguities related
to the scale choice for inclusive jet production. This was achieved by introducing scale choice criteria that make use of arguments based purely on theory and prior to any comparison with experimental data.

\begin{table}[t!]
 \parbox{.45\linewidth}{
 \centering  
  \begin{tabular}{l c c c c}
    \toprule
    & \multicolumn{4}{c}{criterion} \\
    scale & (a) & (b) & (c) & (d) \\
    \midrule
    $p_{T,1}$    & --         & --         & \checkmark & \checkmark \\
    $2\,p_{T,1}$ & \checkmark & --         & \checkmark & \checkmark \\
    $p_{T}$      & --         & \checkmark & \checkmark & \checkmark \\
    $2\,p_{T}$   & \checkmark & \checkmark & \checkmark & \checkmark \\
    $\hat{H}_{T}/2$   & \checkmark & \checkmark & \checkmark & --         \\
    $\hat{H}$     & \checkmark & \checkmark & \checkmark & \checkmark \\
    \bottomrule
  \end{tabular}\\\vspace{0.2cm}(a) $R$=0.7
  }
  \vspace{-0.3cm}
 \hfill
 \parbox{.45\linewidth}{
 \centering
 \begin{tabular}{l c c c c}
    \toprule
    & \multicolumn{3}{c}{criterion} \\
    scale & (a) & (b) & (c) & (d) \\
    \midrule
    $p_{T,1}$    & --         & --         & --         & --           \\
    $2\,p_{T,1}$ & \checkmark & --         & \checkmark & (\checkmark) \\
    $p_{T}$      & --         & --         & --         & --           \\
    $2\,p_{T}$   & \checkmark & \checkmark & \checkmark & \checkmark   \\
    $\hat{H}_{T}/2$   & \checkmark & \checkmark & --         & --           \\
    $\hat{H}_{T}$     & \checkmark & \checkmark & \checkmark & (\checkmark) \\
    \bottomrule
  \end{tabular}\\\vspace{0.2cm}(b) $R$=0.4
 }
 \vspace{0.5cm}
 \caption{Summary of scales vs.\ scale choice criteria for (a) $R$=0.7 and (b) $R$=0.4 cone sizes~\cite{Currie:2018xkj}.}
 \label{tab:scale-criteria}
\end{table}

\section{Phenomenology}
\begin{figure}[t!]
  \begin{subfigure}{.5\textwidth}
  \centering
  \includegraphics[width=5cm]{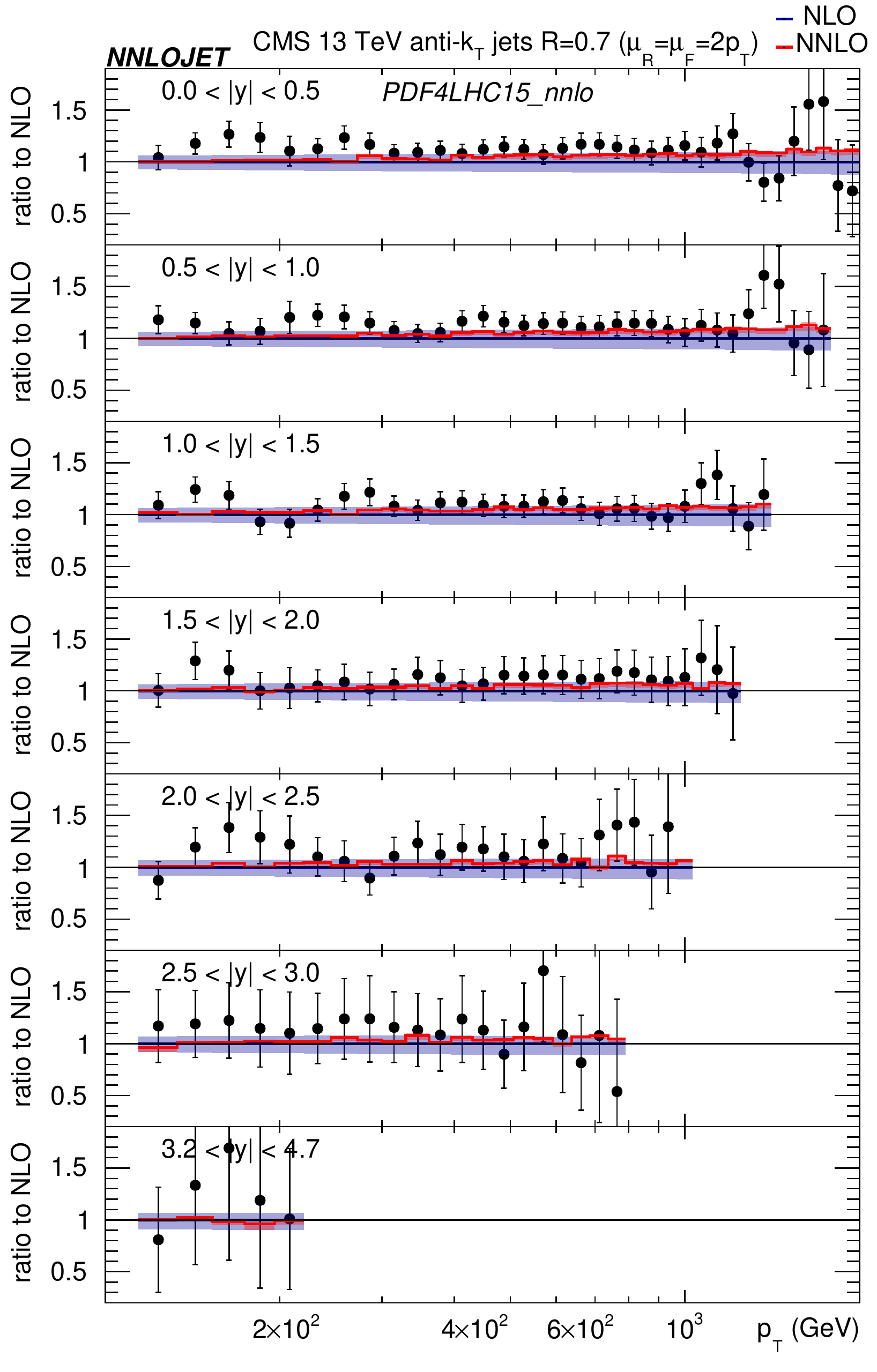}
  \caption{}
  \end{subfigure}%
  \begin{subfigure}{.5\textwidth}
  \centering
  \includegraphics[width=5cm]{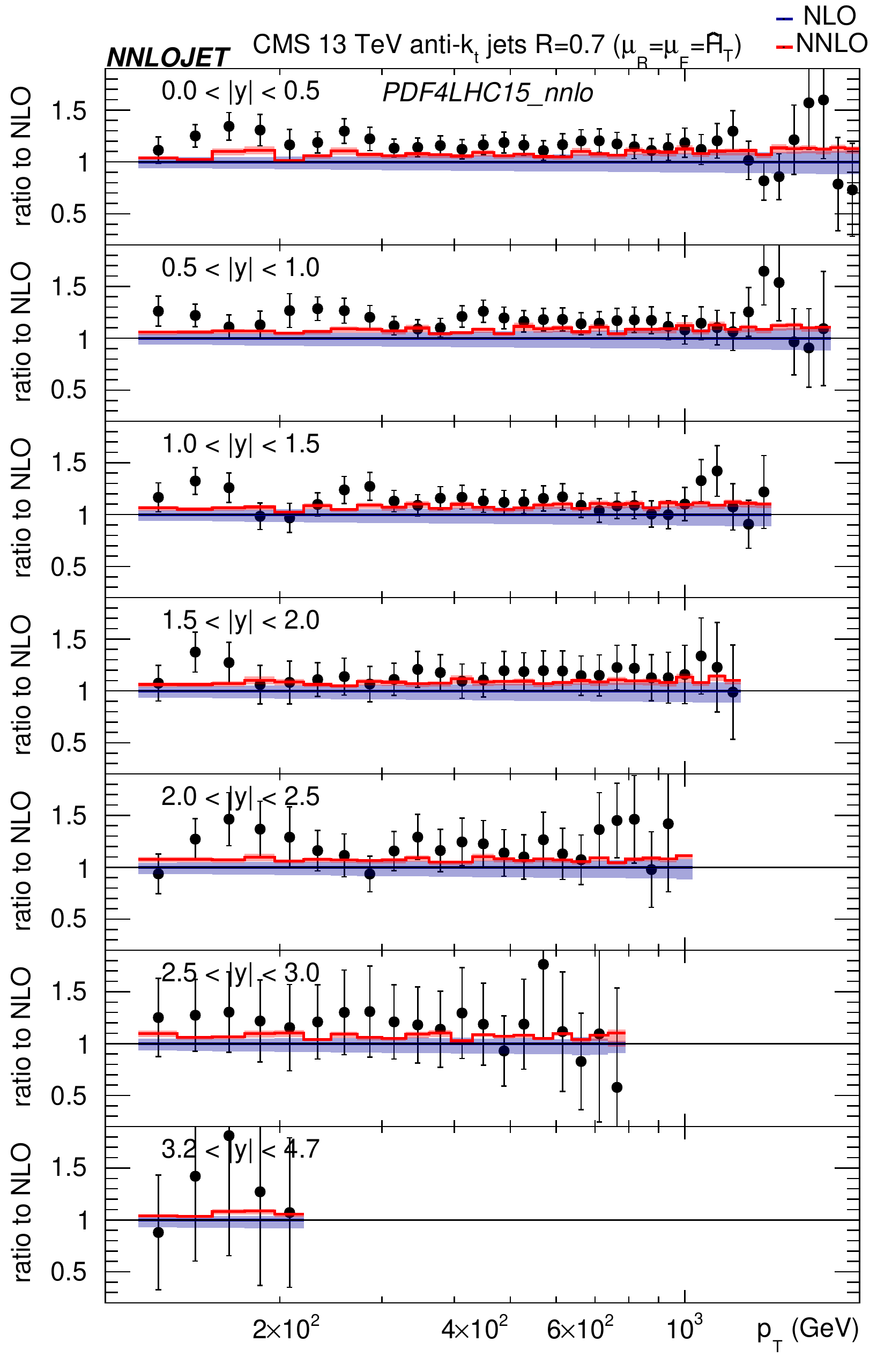}
  \caption{}
  \end{subfigure}  
\caption{Double-differential single jet inclusive cross-sections measurement by CMS~\cite{Khachatryan:2016wdh} and NNLO perturbative QCD predictions as a function of the jet
$p_{T}$ in slices of rapidity, for anti-$k_{T}$ jets with $R=0.7$ normalised to the NLO result for (a) $\mu=2\,p_{T}$, (b) $\mu=\hat{H}_{T}$ scales. 
The shaded bands represent the scale uncertainty. The default PDF set used is {\tt PDF4LHC15nnlo100}~\cite{Currie:2018xkj}.}
\label{fig:pTallscales07}
\end{figure}

Having discussed how the jet kinematics at the LHC differently affects each of the central scale choices,
in this section we now present predictions for the double differential jet cross section at NLO and NNLO for the CMS measurement at $\sqrt{s}=13$ TeV~\cite{Khachatryan:2016wdh}. Our numerical
setup is proton-proton collisions at $\sqrt{s}=13$~TeV for single jet inclusive production where the jets are identified using the anti-$k_T$ algorithm with $R$=0.7. Jets are accepted within the 
fiducial volume defined through the cuts,
\begin{align}
  |y^j| &< 4.7 ,&
  p_T^j &> 114~\textrm{GeV} , 
\end{align}
covering jet-$p_{T}$ values up to $2~$TeV, and ordered in transverse momentum.

Figure~\ref{fig:pTallscales07} displays the NLO and NNLO predictions for the jet-based scale choice $\mu=2 p_{T}$, as well as for the event-based scale choice $\hat{H}_{T}$ compared to the CMS 13 TeV 
data~\cite{Khachatryan:2016wdh} with a jet cone size of $R$= 0.7. For both scale choices we observe small positive NNLO
corrections across all rapidity slices, that improve the agreement with the CMS data,
as compared to the NLO prediction. In addition, we identify a reduction in the scale
uncertainty going from NLO to NNLO across the entire $p_T$ range. We expect that these results will enable precision
phenomenology with jet data, such as the NNLO determination of the parton distributions functions
and of the strong coupling constant from LHC jet data in the near future.

\section*{Acknowledgments}
I would like to thank James Currie, Nigel Glover, Aude Gehrmann-De Ridder, Thomas Gehrmann and Alex Huss for the collaboration on the work reported here.
This work was supported by the Funda\c{c}\~{a}o para a Ci\^{e}ncia e Tecnologia (FCT-Portugal), project UID/FIS/00777/2013.

\end{document}